\documentclass[
superscriptaddress,amsmath,amssymb,aps,longbibliography,twocolumn,showpacs,a4paper]{revtex4-2}

\usepackage{graphicx}
\usepackage{mathtools}
\usepackage{bm}
\usepackage{braket}
\usepackage{amsmath}   
\usepackage{hyperref}  
\usepackage[normalem]{ulem} 
\usepackage{xcolor}         

\newcommand{\redsout}[1]{
  \begingroup
    \color{red}
    \def\UL@color{\relax}
    \sout{#1}
  \endgroup
}
\begin{document}
\title{Benchmarking First-Principles Approaches for Extracting Magnetic Exchange Interactions}
\author{Nafise Rezaei}
\affiliation{Skolkovo Institute of Science and Technology, 121205, Bolshoy Boulevard 30, bld. 1, Moscow, Russia.}
\author{Artem R. Oganov}
\affiliation{Skolkovo Institute of Science and Technology, 121205, Bolshoy Boulevard 30, bld. 1, Moscow, Russia.}
\author{Ali Ghojavand}
\affiliation{COMMIT, Department of Physics, University of Antwerp, Groenenborgerlaan 171, B-2020 Antwerp, Belgium.}
\affiliation{Physique Th\'eorique des Mat\'eriaux, QMAT, Universit\'e de Li\`ege, B-4000 Sart-Tilman, Belgium.}
\author{Milorad V. Milo\v{s}evi\'c}
\affiliation{COMMIT, Department of Physics, University of Antwerp, Groenenborgerlaan 171, B-2020 Antwerp, Belgium.}
\author{Mojtaba Alaei}
\affiliation{Skolkovo Institute of Science and Technology, 121205, Bolshoy Boulevard 30, bld. 1, Moscow, Russia.}
\affiliation{Department of Physics, Isfahan University of Technology, Isfahan 84156-83111, Iran.}
\date{\today}

\begin{abstract}
Magnetic exchange interactions govern the macroscopic magnetic behavior of solids and underpin both fundamental spin phenomena and emerging technologies. The accurate and efficient determination of these interactions is therefore critical for predictive modeling of magnetic materials. Here we present a systematic first-principles comparison of three widely used approaches—the Least-Squares Total Energy (LSTE), the Four-State Total Energy (FSTE), and the Green’s function-based Liechtenstein \textit{et al.} (LKAG) methods—applied to thirteen antiferromagnetic compounds. We introduce an  framework for identifying the minimal supercells required for an accurate exchange parameter extraction in the FSTE method, significantly reducing computational cost while preserving precision. Our results show that LSTE and FSTE yield nearly identical exchange parameters, whereas the LKAG method reproduces the dominant exchange interactions but exhibits quantitative deviations. A detailed analysis of computational efficiency versus accuracy reveals that the LSTE scheme offers the most favorable balance, establishing a general, reproducible, and scalable workflow for Heisenberg mapping, while the FSTE approach remains the most straightforward for extracting specific exchange interactions.
\end{abstract}


\maketitle
\section{\label{sec:int}INTRODUCTION}
Magnetic exchange interactions are fundamental quantum-mechanical quantities that govern the long-range order of magnetic moments in solids. Arising from the interplay between the Pauli exclusion principle and the Coulomb interaction, they underpin the diverse range of magnetic phenomena observed in materials, including ferro-, antiferro-, and ferrimagnetism. 

Accurate determination of these magnetic exchange parameters is essential for understanding and predicting magnetic behavior. Derived typically from first-principles electronic-structure calculations, they serve as key inputs to spin models such as the Heisenberg Hamiltonian, enabling simulations across multiple length and time scales. To address this central challenge, a variety of density functional theory (DFT)-based approaches have been developed, each offering distinct balances between computational cost and accuracy.

A common approach to determine exchange interactions is to evaluate total energy differences among several magnetic configurations obtained from self-consistent first-principles calculations. In a total energy method, the exchange parameters are extracted by mapping the latter energy differences onto a spin Hamiltonian, typically the Heisenberg model. Within this framework, two widely used techniques are the Least-Squares Total Energy (LSTE) method and the Four-State Total Energy (FSTE) method~\cite{4S-0, 4S-1, 4S-2}. Both rely on total energies computed from density functional theory, but differ in how the exchange constants are obtained: the LSTE approach fits multiple spin configurations to minimize deviations from the model, whereas the FSTE method analytically evaluates the interaction between a selected pair of magnetic atoms using a minimal number of configurations.

An alternative route to obtaining exchange interactions derives them directly from the electronic structure of the collinear ground state, thus avoiding multiple total-energy calculations. In the Green’s function formalism developed by Liechtenstein \textit{et al.}~\cite{Green_review}, often referred to as LKAG method \cite{LKAG}, the exchange parameter $J_{ij}$ is evaluated perturbatively from the scattering properties of magnetic sites. Small spin rotations at sites $i$ and $j$ within an otherwise collinear background induce an exchange energy that can be expressed as an energy integral over the spin-resolved Green’s functions, $G^{\uparrow}$ and $G^{\downarrow}$. The difference between these propagators quantifies how a spin fluctuation at one site affects another, and, through the magnetic force theorem, the resulting energy variation is mapped onto the Heisenberg Hamiltonian. The advantage of this method is that exchange interactions can be simultaneously obtained for all pairs of interest, and that the method can be extended to achieve insights even beyond the orbitally-resolved origins of any given exchange interaction \cite{Sabani2025}.

Another recent approach is to compute the magnon dispersion using time-dependent density functional perturbation theory and then extract the exchange interactions by applying linear spin-wave theory to that dispersion~\cite{TDDFT-1}. This procedure mirrors the method used by experimentalists to determine exchange interactions from inelastic neutron-scattering data. Although this approach is quite successful, it is significantly more computationally demanding than the methods discussed above~\cite{TDDFT-2}.

Therefore, in this work, we evaluate the Heisenberg exchange interactions for thirteen antiferromagnetic materials using the three most widely employed first-principles-based methods. We systematically compare their performance in terms of both the magnitude and trends of the computed exchange parameters, as well as the associated computational cost. Our analysis reveals that the choice of computational approach can markedly affect the quantitative and qualitative description of magnetic interactions. By assessing the consistency, strengths, and limitations of each method, this comparative study offers practical guidance for selecting reliable and efficient strategies to investigate magnetism in complex materials.  

The paper is organized as follows. Section~\ref{sec:computational} describes the computational setup used to determine the exchange interactions. Section~\ref{sec:methods} outlines the theoretical and computational frameworks of the LSTE, FSTE, and Green’s function-based LKAG methods, as used in this work. Section~\ref{sec:results} presents and discusses the comparative analysis between the methods. Finally, Section~\ref{sec:conclusion} summarizes our main findings and their implications.

\section{Theoretical and computational methodology}
\subsection{\label{sec:computational}Computational details}
\label{sec:computational}

Most of our first-principles calculations were performed using the VASP package~\cite{vasp}, based on the projector augmented-wave (PAW) method and a plane-wave basis set. A plane-wave energy cutoff of 500~eV was used to ensure convergence of the total energies. For Brillouin zone integration, a $\mathbf{k}$-point mesh with a reciprocal-space resolution of approximately 0.15~\AA$^{-1}$ is employed, generated using the Monkhorst–Pack scheme.

We adopt both the generalized gradient approximation (GGA)~\cite{GGA} in the Perdew–Burke–Ernzerhof (PBE) form~\cite{PBE} and the GGA+$U$ (PBE+$U$) approach~\cite{GGAU1, GGAU2} to account for electron correlation effects in systems with localized $d$ electrons. 

For the Green’s function–based approach of Liechtenstein \textit{et al.}, we utilize the TB2J package~\cite{tb2j}, which interfaces with WANNIER90 ~\cite{wannier90} and SIESTA~\cite{siesta} packages to compute the exchange parameters from the real-space Green’s functions. For the construction of Maximally Localized Wannier Functions (MLWFs), we use atomic-like trial orbitals consisting of $d$ states on the transition-metal sites and $p$ states on the ligand sites. For the calculations with the SIESTA~\cite{siesta} code, we used fully relativistic norm-conserving pseudopotentials from the Pseudo-Dojo database~\cite{siesta-2} in the psml format~\cite{siesta-3}. We employed a mesh-cutoff energy of 400 Ry together with a double-zeta polarized (DZP) basis set for all atoms. 

For all compounds under study we used the experimentally determined crystal structures. The considered materials were the following: Cr$_2$O$_3$~\cite{Cr2O3-S}, Cr$_2$TeO$_6$~\cite{Cr(Fe)2Te(W)O6-S}, MnO~\cite{MnO-S}, MnF$_2$~\cite{MnF2-S}, KMnF$_3$~\cite{KMnF3-S}, KMnSb~\cite{KMnSb-S}, Fe$_2$O$_3$~\cite{Fe2O3-S}, BiFeO$_3$~\cite{BiFeO3-S}, Fe$_2$TeO$_6$~\cite{Cr(Fe)2Te(W)O6-S}, NiO~\cite{NiO-S}, NiF$_2$~\cite{NiF2-S}, NiBr$_2$~\cite{NiBr2-S}, and KNiF$_3$~\cite{KNiF3-S}.

\subsection{Magnetic exchange interactions from first principles}
\label{sec:methods}
\subsubsection{Least-Squares Total Energy (LSTE) method}
A widely used approach for evaluating exchange interactions is the LSTE method, in
which the total energies of multiple magnetic configurations are fitted to a
Heisenberg Hamiltonian. In this framework, the first-principles total energies
$E^{\mathrm{DFT}}$ of different spin arrangements are mapped onto
\[
E^{\mathrm{model}} = -\frac{1}{2} \sum_{i,j} J_{ij}\, \mathbf{S}_i \cdot \mathbf{S}_j ,
\]
where $J_{ij}$ denote the exchange constants to be determined. By constructing
a system of linear equations that relate the calculated DFT energies to the
model expression, the problem can be written as
\[
\mathbf{E}^{\mathrm{DFT}} = \mathbf{A}\, \mathbf{J},
\]
where $\mathbf{E}^{\mathrm{DFT}}$ is the vector of computed total energies,
$\mathbf{A}$ is the coefficient matrix associated with each magnetic
configuration, and $\mathbf{J}$ is the vector of the sought exchange parameters. The exchange constants are then obtained by minimizing the least-squares functional 
$
\chi^2 = \sum_k \big| E_k^{\mathrm{DFT}} - E_k^{\mathrm{model}} \big|^2,$
which leads to the analytical solution \cite{mosleh}
\[
\mathbf{J} = \left( \mathbf{A}^{\mathrm{T}} \mathbf{A} \right)^{-1}
\mathbf{A}^{\mathrm{T}} \mathbf{E}^{\mathrm{DFT}}.
\]
This procedure yields the set of $J_{ij}$ values that best reproduce the
relative DFT energies. The reliability of the extracted parameters depends
sensitively on the chosen magnetic configurations and the number of
linearly independent equations.

In the LSTE approach, determining \( n \) independent exchange parameters requires constructing at least \( n + 1 \) linearly independent total-energy equations corresponding to distinct magnetic configurations. To achieve this, one must employ a suitable supercell.  
A systematic procedure for supercell selection has recently been proposed and implemented in the SUPERHEX code~\cite{superhex}.
In this framework, supercells are generated using hermite normal form matrices, and the null space of the Heisenberg Hamiltonian coefficient matrix (\( \mathbf{A} \)) is analyzed to identify the maximum number of exchange parameters that can be uniquely determined for each supercell. 
In the present work, we employ the smallest suitable supercell identified through this analysis to compute the exchange interactions within the LSTE method. Subsequently, a series of magnetic configurations is generated randomly, and their number is gradually increased until the calculated exchange parameters reach convergence.

\subsubsection{Four-State Total Energy (FSTE) method}
A more direct approach to computing the exchange interactions is the FSTE method~\cite{4S-0, 4S-1, 4S-2}. It comprises evaluating the total energies of four distinct spin configurations ---combinations of parallel and antiparallel alignments of the target spin pair, while keeping all other spins fixed. 
Suppose that our objective is to evaluate the exchange interaction between a selected pair of sites $(i,j)$. In the FSTE method, 
the total energies are computed for four distinct collinear spin configurations in which 
only the orientations of spins at $i$ and $j$ are reversed, while the magnetic moments 
of all other sites are kept fixed.
Denote the scalar spin variables \(s_k=\pm1\) for the direction of \(\mathbf{S}_k\) and \(S\) for the spin magnitude so that \(\mathbf{S}_k = s_k S\,\hat{z}\). For brevity define
\[
A \equiv \sum_{k\neq i,j} J_{i k}\, s_k,\qquad
B \equiv \sum_{k\neq i,j} J_{j k}\, s_k,
\]
and let \(C\) collect all contributions that do not depend on \(s_i\) or \(s_j\) (i.e. interactions among the other sites and constant terms).

The total energy for a general choice \((s_i,s_j)\) can be written as
\[
E^{s_i s_j}
=-m_{ij}J_{ij} \, S^2 \, s_i s_j  - S^2\, s_i A - S^2\, s_j B + C,
\]
where $m_{ij}$ is the multiplicity, i.e., the number of equivalent bonds associated with this atomic pair within the supercell, including their periodic images.
 Now insert the four collinear configurations \( (\uparrow,\uparrow)=(+1,+1)\), \((\downarrow,\downarrow)=(-1,-1)\),
\((\uparrow,\downarrow)=(+1,-1)\) and \((\downarrow,\uparrow)=(-1,+1)\). This yields
\[
E^{\uparrow\uparrow} 
=-m_{ij}J_{ij}S^2 - S^2 A - S^2 B + C,\]
\[E^{\downarrow\downarrow} =
 -m_{ij}J_{ij}S^2 + S^2 A + S^2 B + C,\]
\[
E^{\uparrow\downarrow} =
 m_{ij}J_{ij}S^2 - S^2 A + S^2 B + C,\]
and
\[ E^{\downarrow\uparrow} 
= m_{ij}J_{ij}S^2 + S^2 A - S^2 B + C.
\]
Therefore the exchange parameter can then be extracted as
\[
J_{ij} = \frac{1}{4m_{ij}S^2} 
\Big( E^{\uparrow\downarrow} + E^{\downarrow\uparrow} - E^{\uparrow\uparrow} - E^{\downarrow\downarrow} \Big).
\]
Thus, the FSTE method provides a simple and robust scheme for computing exchange
interactions directly from total energy calculations.

For the calculation of exchange interactions within the FSTE method, a sufficiently large supercell is required. In practice, however, this aspect is often treated heuristically, and oversized supercells are chosen without a clear justification. Here, we explicitly formulate the criteria that a supercell must satisfy for reliable FSTE calculations and introduce an efficient procedure to determine the optimal one.  

Specifically, the supercell must ensure that, for every magnetic-atom pair within the \( n \)-th neighbor distance (where \( n \) denotes the maximum interaction range considered), no periodic image of either atom appears at a distance shorter than \( d_n \). This condition eliminates artificial coupling between periodic replicas and guarantees that the extracted exchange parameters reflect only the intended spin interactions.

To identify the optimal supercell that satisfies the conditions described above, we have extended the capabilities of the SUPERHEX code. 
The procedure begins by generating a set of candidate supercells using SUPERHEX. 
Each supercell is then analyzed to determine which exchange interactions (\( J_1, J_2, \dots, J_n \)) can be uniquely and independently extracted within its geometry, as well as to identify the specific spin pairs whose reversal enables the calculation of each exchange parameter. 
In our calculations, the smallest supercell capable of capturing all desired exchange interactions is selected, thereby minimizing computational cost while maintaining high accuracy. For all compounds considered in this work, the FSTE calculations are performed starting from the ferromagnetic configuration as the initial magnetic state.

\begin{figure*}
    \centering
    \includegraphics[width=\linewidth]{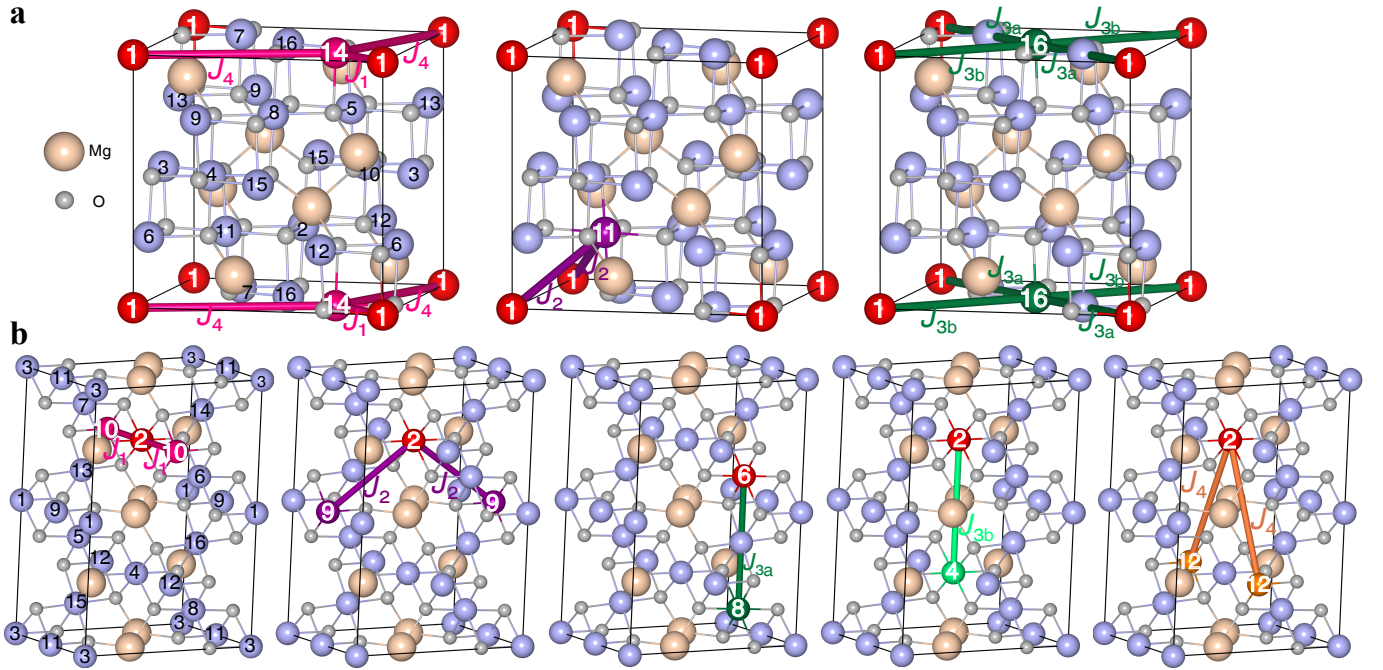}
\caption{
Crystal structure of MgCr$_2$O$_4$, with Cr atoms labeled by numbers. 
(a) The conventional cell, in which only the $J_2$ exchange interaction can be calculated. Other interactions cannot be determined in this cell because, for example, if one takes (Cr1, Cr14) as a first-neighbor pair, its periodic image (Cr1+image, Cr14) appears as a fourth neighbor as well. Similar occurs for the (Cr1, Cr16) pair, preventing calculation of the $J_{3a}$ and $J_{3b}$ interactions. 
(b) The supercell generated and selected using the updated SUPERHEX code, which enables the calculation of exchange interactions up to the fourth-nearest neighbors.
The Cr–Cr pairs whose spin reversals are employed to extract the corresponding exchange parameters are highlighted.
\label{fig:1}
}
\end{figure*}
As an illustration, we consider the MgCr$_2$O$_4$ compound, which was the first system used to demonstrate the FSTE method~\cite{4S-0}. 
In that study, a large \(2 \times 2 \times 2\) supercell of the conventional cell (containing 248 atoms) was employed to calculate four exchange interactions, requiring substantial computational resources. 
In contrast, using the method introduced above, we have identified a supercell containing only 56 atoms that is capable of resolving the exchange interactions up to the fourth nearest neighbor, 
whereas the conventional MgCr$_2$O$_4$ cell with the same number of atoms (56 atoms) allows the calculation of only the \(J_2\) exchange interaction (see Fig.~\ref{fig:1}).

We evaluate the exchange interactions of MgCr$_2$O$_4$ using both the large \(2 \times 2 \times 2\) supercell and the optimized 56-atom supercell obtained from the SUPERHEX analysis, in order to assess the reliability of the results with respect to the chosen cell size. 
All calculations are performed within the GGA+$U$ framework, with on-site Coulomb and exchange parameters of \(U = 3~\text{eV}\) and \(J = 0.9~\text{eV}\), respectively, as adopted from Ref.~\cite{4S-0}. 
A plane-wave energy cutoff of 400~eV is employed in all cases, and the experimental crystal structure is used without atomic relaxation~\cite{mgcr2o4}. 
For the large \(2 \times 2 \times 2\) supercell, the extracted exchange interactions are 
\(J_1 = -8.80~\text{meV}\), \(J_2 = 0.03~\text{meV}\), \(J_{3a} = -0.63~\text{meV}\), \(J_{3b} = -0.20~\text{meV}\), and \(J_4 = 0.05~\text{meV}\). 
These values are in good agreement with those obtained using the smaller supercell generated by the SUPERHEX code, 
which yields \(J_1 = -8.67~\text{meV}\), \(J_2 = 0.02~\text{meV}\), \(J_{3a} = -0.44~\text{meV}\), \(J_{3b} = -0.23~\text{meV}\), and \(J_4 = 0.05~\text{meV}\), assuming \(S = 1\).

\subsubsection{Green's functions-based LKAG method}
The variation of the DFT total energy due to small changes in the directions of the magnetic moments at sites $i$ and $j$ ($\delta \mathbf{S}_i$ and $\delta \mathbf{S}_j$) can be related to the exchange interaction parameter $J_{ij}$ as~\cite{Green_review, tb2j}:
\[
\delta E_{ij} = -J_{ij}\, \delta \mathbf{S}_i \cdot \delta \mathbf{S}_j,
\]
where $\mathbf{S}$ is the spin moment.  
Within the magnetic force theorem, the total energy variation can be approximated by the change in the sum of occupied single-particle eigenvalues, evaluated within the fixed ground-state potential:
\[
\delta E_{ij} \approx \int_{-\infty}^{E_F} \epsilon\, \delta n(\epsilon)\, d\epsilon,
\]
where $n(\epsilon)$ is the electronic density of states and $E_F$ is the Fermi energy.  

To evaluate $\delta E_{ij}$ explicitly, one can employ the Green’s function formalism, in which the energy change associated with infinitesimal spin rotations is obtained from the second-order perturbation theory~\cite{Green_review, Green_review2,Korotin2015}:
\[
\delta E_{ij} = -\frac{2}{\pi} \int_{-\infty}^{E_F} 
\mathrm{Im}\, \mathrm{Tr}
\!\left(
\delta \mathbf{H}_i\, \mathbf{G}\,
\delta \mathbf{H}_j\, \mathbf{G}
\right)
d\epsilon,
\]
where $\delta \mathbf{H}_i$ is the perturbation to the Hamiltonian due to the rotation of the magnetic moment at site $i$, and $\mathbf{G}$ is the Green’s function of the unperturbed (collinear) reference state.  

Replacing $\delta \mathbf{H}_i$ by the local exchange splitting 
$\mathbf{\Delta}_i = H_{ii}^{\uparrow} - H_{ii}^{\downarrow}$ 
leads to the expression for the exchange interaction~\cite{Green_review,tb2j}:
\[
J_{ij} = -\frac{1}{2\pi} \int_{-\infty}^{E_F} 
\mathrm{Im}\, \mathrm{Tr}
\!\left(
\mathbf{\Delta}_i\, \mathbf{G}_{ij}^{\uparrow}\,
\mathbf{\Delta}_j\, \mathbf{G}_{ji}^{\downarrow}
\right)
d\epsilon.
\]

Thus, the Green’s function-based approach of Liechtenstein \textit{et al.} provides a direct and efficient means of evaluating the energy variation $\delta E$ associated with infinitesimal spin rotations, allowing the exchange parameters $J_{ij}$ to be extracted from first-principles electronic structure calculations.  
Furthermore, because all quantities are expressed in a local orbital basis, this method naturally enables an orbital-resolved decomposition of $J_{ij}$, offering microscopic insight into the mechanisms of magnetic coupling.

The calculation of exchange interactions using the Green's function-based LKAG method, as implemented in the TB2J package, can be performed directly on the ground-state primitive cell, without the need for supercells. In this approach, the electronic Green's function is required, which depends on a Hamiltonian expressed in a tight-binding basis. Such a Hamiltonian is typically constructed either from MLWFs using WANNIER90, or directly from localized atomic orbitals provided by codes such as SIESTA. 
Consequently, the accuracy of the computed exchange interactions depends on the quality of the underlying localized basis; in particular, for the MLWF approach, the results are sensitive to the quality of the Wannierization procedure.

\section{\label{sec:results}Comparative analysis of the magnetic exchange interactions}



\begin{table}[!]
\centering
\caption{Pearson correlation coefficients between the exchange parameters $J_n$
obtained from different methods for each compound within the GGA+$U$ approximation.}
\label{tab:pearson}
\renewcommand{\arraystretch}{1.15}
{\footnotesize
\begin{tabular}{lcccc}
\hline
\footnotesize \textbf{Compound} & \footnotesize$U$(eV)& \footnotesize(FSTE/LSTE) & \footnotesize(TB2J/LSTE) & \footnotesize(TB2J/FSTE) \\
\hline
Cr$_2$O$_3$   & 2.38 &  0.999 & 0.997 & 0.998 \\
Cr$_2$TeO$_6$ & 3.47 &  1.000 & 0.998 & 0.998 \\
MnO           & 3.43 &  1.000 & 0.886 & 0.899 \\
MnF$_2$       & 3.39 &  1.000 & 0.993 & 0.993 \\
KMnF$_3$      & 3.08 & 1.000 & 1.000 & 1.000 \\
KMnSb         & 2.93 & 1.000 & 1.000 & 1.000 \\
Fe$_2$O$_3$   & 4.69 & 0.994 & 0.994 & 0.989 \\
BiFeO$_3$     & 4.59 & 1.000 & 1.000 & 0.999 \\
Fe$_2$TeO$_6$ & 5.33 & 1.000 & 0.999 & 1.000 \\
NiO           & 4.85 & 1.000 & 0.997 & 0.998 \\
NiF$_2$       & 5.14 & 0.999 & 0.984 & 0.987 \\
NiBr$_2$      & 6.35 & 1.000 & 0.750 & 0.755 \\
KNiF$_3$      & 4.65 & 1.000 & 1.000 & 1.000 \\
\hline
\end{tabular}
}
\end{table}

Determining the on-site Coulomb parameter ($U$) is essential for the GGA+$U$ approach. 
Although the aim of this study is not to obtain $U$ values optimized for a specific property, such as the transition temperature, 
a consistent and practical procedure is applied to all compounds. To this end, hybrid functional (HSE06~\cite{HSE06}) calculations are performed to evaluate the energy difference between the antiferromagnetic and ferromagnetic configurations. Subsequently, \( U \) is varied within the GGA+\( U \) scheme, and the value that best reproduces the HSE06 energy difference between these two magnetic states is selected as the optimal \( U \). The resulting \( U \) values for all compounds are listed in Table~\ref{tab:pearson}.

After determining the $U$ parameters, the exchange interactions were calculated using the 
LSTE, FSTE, and TB2J methods, each both within the GGA and GGA+$U$ frameworks. 
All calculated magnetic exchange interactions are reported in the Supplementary Material~\cite{SM}. In our calculations, we assume $S=1$. The results are compared from three perspectives: 
(1) the precision and numerical stability of each method, 
(2) the relative behavior and consistency of the exchange interactions among the methods, 
and (3) the computational cost and efficiency of each approach.

\begin{figure}
    \centering
    \includegraphics[width=1\linewidth]{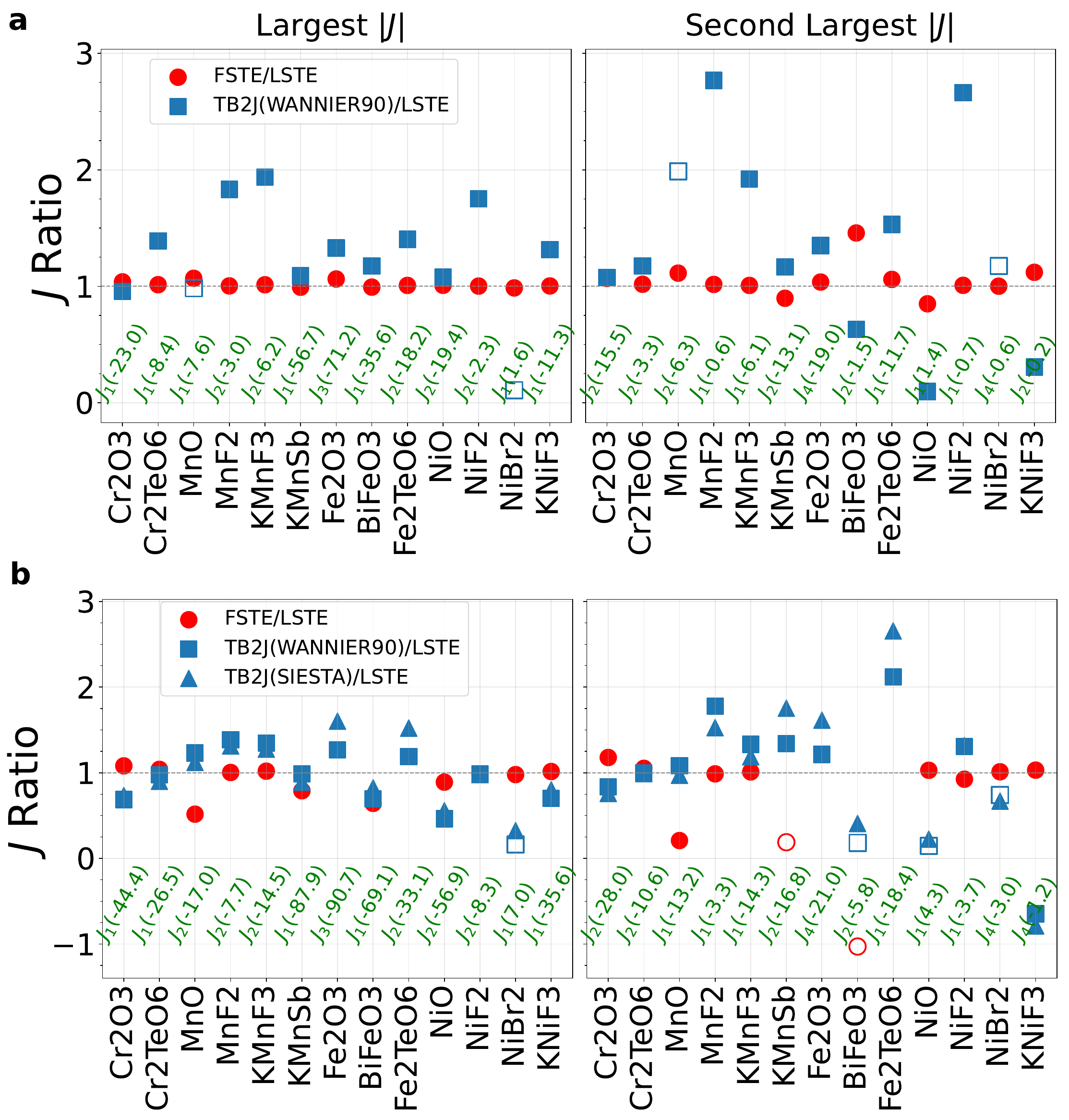}
    \caption{Comparison of the exchange parameter ratios relative to the LSTE method, for the largest and the second-largest $J$ values, in (a) GGA+$U$ (b) GGA approximation. Filled symbols: same $J$ interaction identified as top coupling in both methods. Empty symbols: different $J$ interactions ranked as top couplings. $J$ annotations show the specific exchange interaction being scaled for each data point, in meV.}
    \label{fig:J-ratio}
\end{figure}

For each compound, we compare the ratios of the largest and second-largest magnetic exchange interactions obtained from the FSTE and TB2J methods to those derived from the LSTE approach. 
The results for both GGA and GGA+$U$ approximations are presented in Fig.~\ref{fig:J-ratio}. 
As shown in Fig.~\ref{fig:J-ratio}(a), the ratios of the FSTE to LSTE exchange interactions for the dominant coupling (\(J_{\text{max}}\)) 
are generally close to unity within the GGA+$U$ approximation, indicating excellent agreement between the two methods. 
For the second-largest interaction, the ratios also remain close to unity for most compounds, except for BiFeO\(_3\), where a moderately deviating value of approximately 1.5 is obtained. 
However, since the magnitude of the second-largest coupling in BiFeO\(_3\) is significantly smaller than the dominant one, 
this deviation is negligible in practice. 
In Fig.~\ref{fig:J-ratio}, filled symbols denote cases where the FSTE or TB2J methods identify the same leading exchange interaction (i.e., the same pair of magnetic ions with the largest $|J|$ value) as the LSTE method, whereas open symbols denote cases where a different exchange interaction is ranked as dominant.
Overall, the FSTE and LSTE approaches exhibit strong consistency in predicting the hierarchy of exchange interactions across all studied compounds.

Figure~\ref{fig:J-ratio} (a) also presents the ratios of exchange interactions obtained from the TB2J using the WANNIER90 interface 
with respect to those from the LSTE method in GGA+$U$ framework. 
For the largest exchange interaction, this ratio varies between 0.1 and 1.93, whereas for the second-largest exchange interaction it ranges from 0.1 to 2.77.
In some compounds, the TB2J (WANNIER90) results demonstrate good agreement with the LSTE method (shown ratio close to unity), while in other cases TB2J predicts exchange interactions up to twice larger than those obtained from LSTE. 
Part of the discrepancy between the total-energy methods and the Green’s function–based LKAG method originates from the magnetic force theorem, which assumes that small changes in the directions of magnetic moments do not affect the charge density but only the band energies. 
Therefore, only the perturbative variations in band energies caused by changes in the magnetic moment orientations are considered. 
Recent studies~\cite{FT_problem1, FT_problem2, TDDFT-2} have demonstrated noticeable differences between fully self-consistent calculations and those based on the magnetic force theorem, indicating that in some cases these deviations can be significant. 
Surprisingly, the TB2J code 
yields two inequivalent first-neighbor exchange parameters for symmetry-equivalent Ni–Ni (Mn-Mn) bonds in NiO and MnO, even though the structures are fully symmetric and free of tilting or rhombohedral distortion. In the presence of rhombohedral distortion, such a splitting is expected because slight differences in bond lengths lead to distinct exchange interactions. However, in the undistorted structures all first-neighbor bonds are symmetry-equivalent, and therefore their exchange parameters should have a unique value \cite{Jacobsson2013}.
In Fig.~\ref{fig:J-ratio}, the larger of the two values is plotted.

Despite these quantitative differences, the TB2J(WANNIER90) approach generally identifies the same dominant exchange interactions 
as the LSTE and FSTE methods. The Pearson correlation coefficients between the exchange interactions obtained from the three methods within the GGA+$U$ framework 
are summarized in Table~\ref{tab:pearson}. 
Once again, the values close to unity indicate a strong consistency between methods. 
Overall, the high correlation values demonstrate that all three approaches predict the relative ordering of exchange interactions consistently. 
Nevertheless, some deviations are observed between the TB2J results and those from the LSTE and FSTE methods.


In Fig.~\ref{fig:J-ratio}(b), the results for the largest and second-largest exchange interactions obtained within the GGA approximation are shown. For Fe$_2$TeO$_6$ and Fe$_2$O$_3$, the FSTE calculations could not be converged due to difficulties in stabilizing the required magnetic configurations. 
In the case of Fe$_2$TeO$_6$, we employed the larger supercell used in the GGA+$U$ calculations to enable access to a broader set of magnetic configurations within the LSTE method.

Overall, the largest and second-largest exchange interactions obtained from the FSTE method relative to the LSTE method remain close to unity, indicating strong consistency between the two approaches. Results for a few compounds, however, exhibit larger deviations between the methods. Closer examination reveals that these systems often display a more metallic character in certain magnetic configurations, which can affect the accuracy of total-energy calculations and, consequently, the extracted exchange parameters.

A notable case is BiFeO$_3$, where a significant difference is observed between the two methods: 
$J_1^{\text{LSTE}} = -69.1$~meV and $J_1^{\text{FSTE}} = -44.4$~meV. 
In the LSTE method, the exchange interactions are calculated using a supercell containing 40 atoms, whereas in the FSTE calculations, a 50-atom supercell is used. To clarify the origin of this discrepancy, we recalculated the exchange interactions within the LSTE method using the same 50-atom supercell as in the FSTE method. 
For this supercell, we computed the total energies for all possible distinct magnetic configurations.

\begin{figure}
    \centering
    \includegraphics[width=0.9\linewidth]{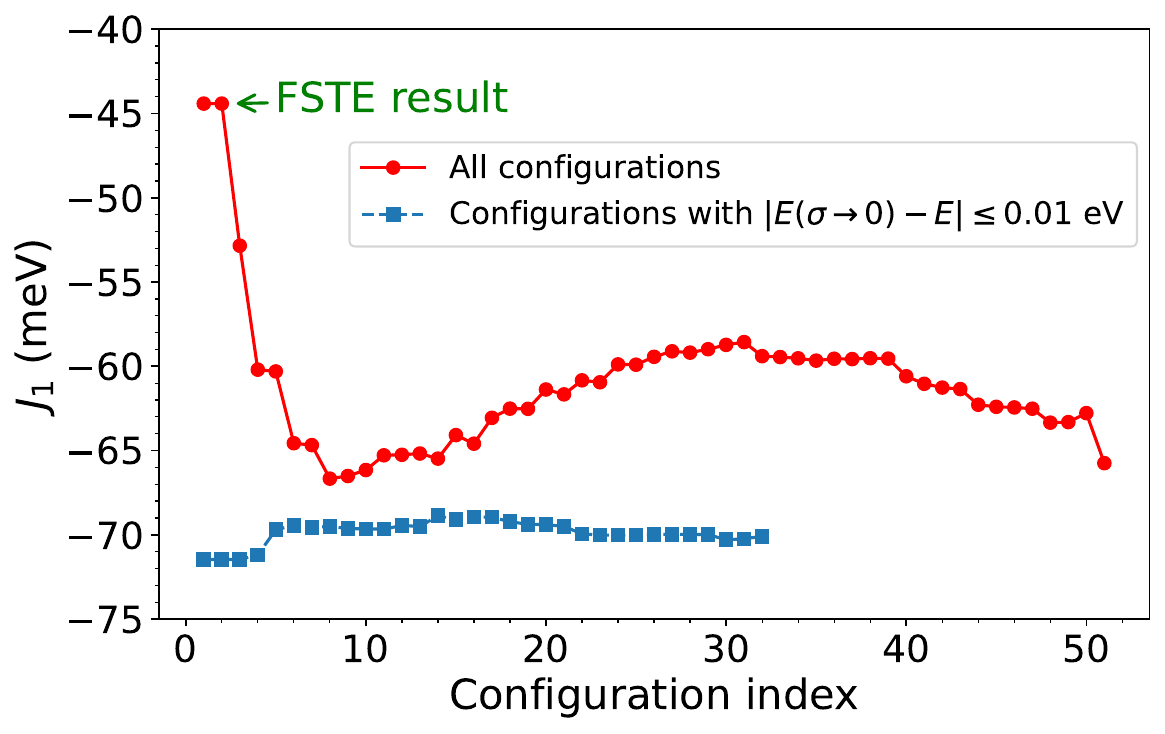}
\caption{
Convergence behavior of the \( J_1 \) exchange parameter in BiFeO\(_3\), using LSTE method within the GGA approximation in a 50-atom supercell, as a function of the number of magnetic configurations. Dots represent results obtained using all possible magnetic configurations, while square symbols correspond to the data excluding configurations with \(|E(\sigma \rightarrow 0) - E| > 0.01~\text{eV}\). Here, \(\sigma\) denotes the electronic smearing width used in the total energy calculations.
\label{fig:J1-4SvsLS}}
\end{figure}

Figure~\ref{fig:J1-4SvsLS} illustrates the convergence behavior of the largest exchange interaction ($J_1$) 
as a function of the number of included magnetic configurations. 
Two cases are shown: in the first, all magnetic configurations are included, while in the second, 
we exclude configurations exhibiting stronger metallic behavior, identified by 
$\lvert E(\sigma \rightarrow 0) - E \rvert > 0.01$~eV. 
Here, $E(\sigma \rightarrow 0)$ denotes the extrapolated total energy in the limit of vanishing electronic smearing 
($\sigma \rightarrow 0$), which corresponds to the ground-state electronic energy without the entropic correction arising from partial occupancies. 
A large difference between $E(\sigma \rightarrow 0)$ and $E$ indicates a significant contribution from electronic entropy, 
implying that the system exhibits partial metallic character in that configuration. 
When all configurations are considered, the first ten correspond exactly to those used in the FSTE method. 
As seen in the figure, the $J_1$ value initially matches the FSTE result but gradually decreases as more configurations are included, 
eventually converging to the value obtained from the restricted (more restrictively insulating) set of configurations. 

We further examined the sensitivity of the FSTE-derived exchange interactions to the initial magnetic state. 
For this purpose, we started from the antiferromagnetic configuration and modified the magnetic pairs of BiFeO$_3$ within the GGA framework. The resulting exchange interactions are 
$J_1 = -71.39$~meV, $J_2 = -10.37$~meV, $J_3 = -4.48$~meV, and $J_4 = -3.62$~meV, 
in stark contrast to the values obtained when starting from the ferromagnetic configuration, 
$J_1 = -44.41$~meV, $J_2 = 5.99$~meV, $J_3 = 7.10$~meV, and $J_4 = 6.34$~meV. 
The exchange interactions obtained from the antiferromagnetic starting configuration are found to be very close 
to those derived from the LSTE method. 
These results demonstrate noticeable differences between the two sets of parameters, 
revealing a clear sensitivity of the FSTE-derived exchange interactions to the chosen initial magnetic configuration.
We also investigated the same effect within the GGA+$U$ framework for the MnO compound. 
The obtained exchange parameters are $J_1 = -7.36$~meV, $J_2 = -6.71$~meV, $J_3 = -0.19$~meV, and $J_4 = -0.28$~meV when starting from the antiferromagnetic configuration, 
compared to $J_1 = -8.12$~meV, $J_2 = -6.98$~meV, $J_3 = -0.06$~meV, and $J_4 = -0.46$~meV when the ferromagnetic initial configuration is used. In this framework, the FSTE-derived exchange interactions are clearly less sensitive to the chosen initial magnetic configuration, but moderate differences remain.

The TB2J results obtained in the GGA framework, similar to those using the GGA+$U$ approximation, exhibit deviations from the LSTE and FSTE methods.
To evaluate the sensitivity of the TB2J outcomes to the Wannierization procedure, we additionally computed the exchange interactions using the TB2J interface to the SIESTA package within the same GGA framework. 
As shown in Fig.~\ref{fig:J-ratio}(b), the dominant exchange interactions derived from TB2J(SIESTA) closely match those obtained from TB2J(WANNIER90), although moderate differences are observed for the second-largest couplings in some compounds. Importantly, the leading exchange interactions identified by TB2J(SIESTA) are fully consistent with those from the LSTE method, whereas TB2J(WANNIER90) exhibits discrepancies in two cases.

\begin{table}[!]
\centering
\caption{Structural and computational parameters for each compound, when studied by LSTE and FSTE GGA+U methods. $N_J$ denotes the number of exchange interactions calculated for each compound. $N_{\mathrm{atoms}}^{\mathrm{method}}$ is the number of atoms in the supercell used by each method, and $N_{\mathrm{runs}}^{\mathrm{method}}$ is the number of self-consistent runs required to compute the $N_J$ exchange interactions.  The computational cost ratio is estimated as $(N_{\mathrm{atoms}}^{\mathrm{FSTE}} / N_{\mathrm{atoms}}^{\mathrm{LSTE}})^3 \times (N_{\mathrm{runs}}^{\mathrm{FSTE}} / N_{\mathrm{runs}}^{\mathrm{LSTE}})$.}
\label{tab:N}
\renewcommand{\arraystretch}{1.15}
\begin{tabular}{lcccccc}
\hline
\textbf{Compound} &
$N_J$ &
$N_{\text{atoms}}^{\text{LSTE}}$ &
$N_{\text{atoms}}^{\text{FSTE}}$ &
$N_{\text{runs}}^{\text{LSTE}}$ &
$N_{\text{runs}}^{\text{FSTE}}$ &
Cost ratio \\
\hline
Cr$_2$O$_3$    & 9 & 40 & 80  & 27 & 20 &  5.93 \\
Cr$_2$TeO$_6$  & 5 & 36 & 36  & 15 & 12 &  0.80 \\
MnO            & 4 & 16 & 32  & 12 & 10 &  6.63 \\
MnF$_2$        & 5 & 36 & 36  & 15 & 12 &  0.80 \\
KMnF$_3$       & 9 & 80 & 160 & 27 & 20 &  5.93 \\
KMnSb          & 6 & 36 & 48  & 23 & 14 &  1.44 \\
Fe$_2$O$_3$    & 9 & 40 & 60  & 27 & 20 &  2.50 \\
BiFeO$_3$      & 4 & 40 & 50  & 12 & 10 &  1.63 \\
Fe$_2$TeO$_6$  & 5 & 36 & 36  & 15 & 12 &  0.80 \\
NiO            & 4 & 16 & 32  & 12 & 10 &  6.67 \\
NiF$_2$        & 5 & 36 & 36  & 15 & 12 &  0.80 \\
NiBr$_2$       & 6 & 54 & 72  & 18 & 14 &  1.84 \\
KNiF$_3$       & 6 & 60 & 120 & 18 & 14 &  6.22 \\
\hline
\end{tabular}
\end{table}

Finally, we compare the computational cost of the total energy methods. 
The TB2J approach is excluded from this comparison not because it showed occasional qualitative differences in results but because it requires only a single DFT calculation on the primitive cell to extract the full set of exchange interactions, so its computational expense cannot be compared on equal footing with the supercell-based FSTE and LSTE methods.
Table~\ref{tab:N} summarizes the number of atoms in the supercells used for the FSTE and LSTE methods, 
as well as the number of total-energy calculations required to obtain the desired exchange interactions. 
For the LSTE method, the number of runs corresponds to the number of spin configurations necessary to achieve 
converged exchange interactions. In most cases, less than three times the number of exchange interactions is sufficient to reach convergence; however, we report three times this number in the table for consistency, except for the KMnSb compound, which requires larger number of runs. 
In the FSTE method, we select the magnetic pairs involved in all interactions that share the same atomic sites, 
so that $E^{\uparrow\uparrow}$ and $E^{\downarrow\uparrow}$ are identical for all exchange interactions. 
Consequently, $2\times (N_J + 1)$ total-energy calculations are required for each compound.
All data are presented within the GGA+$U$ framework. 
The last column of Table~\ref{tab:N} reports the estimated computational cost, which is defined as
\[
\text{Cost ratio} = 
(\frac{N_{\mathrm{atoms}}^{\mathrm{FSTE}}}{N_{\mathrm{atoms}}^{\mathrm{LSTE}}})^3 
\times \frac{N_{\mathrm{runs}}^{\mathrm{FSTE}}}{N_{\mathrm{runs}}^{\mathrm{LSTE}}}.
\]
As shown in Table~\ref{tab:N}, in most cases the supercells required for the FSTE calculations are larger than those used in the LSTE method, 
while the number of magnetic configuration runs is slightly smaller for FSTE. 
Consequently, the cost ratio is generally greater than one, indicating that the LSTE method is computationally more efficient. 
This is primarily due to the cubic scaling of the computational cost with the number of atoms, which dominates over the reduction in the number of runs. Therefore, in most cases, the LSTE approach is significantly less computationally demanding than the FSTE method.

\section{\label{sec:conclusion}CONCLUSIONS}
In this work, we systematically compared three approaches—the LSTE, FSTE, and the Green’s function-based LKAG method as implemented in TB2J—for evaluating magnetic exchange interactions in antiferromagnetic materials. Across a series of transition-metal compounds, both LSTE and FSTE methods produced nearly identical dominant exchange parameters, with strong Pearson correlations. The TB2J results based on VASP+WANNIER90 reproduced the qualitative trends of the exchange couplings, although quantitative deviations up to a factor of two were observed for certain materials.

We also introduced and implemented in the SUPERHEX code an automated algorithm to determine the minimal supercell required for the FSTE method, providing a general and efficient route for extracting Heisenberg exchange parameters. A detailed cost analysis revealed that the LSTE scheme offers computational advantage due to its smaller supercell size and the cubic scaling of computational expense with system size. Nevertheless, the FSTE method remains a practical and transparent option when specific exchange interactions are of interest. 
We note however that convergence failures in certain configurations, and observed dependence on the initial magnetic configuration, can render the FSTE approach unreliable, as it depends on complete convergence of all target magnetic states. In contrast, the LSTE framework allows for the exclusion of problematic cases while still yielding stable and accurate exchange parameters from the remaining data. Overall, our comparative analysis clarifies the strengths and limitations of these widely used approaches and offers practical guidance for selecting reliable and efficient strategies to evaluate magnetic interactions in complex materials.

\section*{Acknowledgements}
This work was supported by the Russian Science
Foundation (Grant 19-72-30043), and the FWO-FNRS Excellence of Science project Shape-ME.
\bibliography{ref.bib}
\end{document}